\documentclass{article}

\usepackage{graphicx} 
\usepackage{authblk}
\usepackage{hyperref}  
\usepackage{amsmath}
\usepackage{tikz}
\usepackage{color}
\usepackage[separate-uncertainty=true]{siunitx}
\usepackage{booktabs}
\usetikzlibrary{matrix}
\usepackage{multirow}
\usepackage{ulem}

\newcommand{\WmK}{W~m$^{-1}$~K$^{-1}$}

\begin{document}

\title{Calibration of Scanning Thermal Microscope using Optimal Estimation of Function Parameters by Iterated Linearization}

\author[1]{Anna Charvátová Campbell \footnote{Corresponding author: anna.charvatovacampbell@cmi.gov.cz}
}

\author[1]{Petr Klapetek}

\author[1]{Radek Šlesinger}

\author[1]{Jan Martinek}

\author[1]{Václav Hortvík}

\author[2]{Viktor Witkovský}

\affil[1]{Czech Metrology Institute,
Okružní 31, 638\,00,
Brno, Czech Republic}
\affil[2]{Institute of Measurement Science, Slovak Academy of Sciences, Dúbravská cesta~9, 841\,04, Bratislava, Slovakia} 

\author[3]{Gejza Wimmer}
\affil[3]{Mathematical Institute, Slovak Academy of Sciences, Štefánikova~49, 814\,73, Bratislava, Slovakia}


    \maketitle
\begin{abstract}
    Scanning thermal microscopy is a unique tool for the study of thermal properties at the nanoscale. However, calibration of the method is a crucial problem. When analyzing local thermal conductivity, direct calibration is not possible and reference samples are used instead. As the calibration dependence is non-linear and there are only a few calibration points, this represents a metrological challenge that needs complex data processing. In this contribution we present use of the OEFPIL algorithm for robust and single-step evaluation of local thermal conductivities and their uncertainties, simplifying this procedure. Furthermore, we test the suitability of SThM calibration for automated measurement.\\
    \noindent
    \textbf{Keywords:} SThM calibration; uncertainty analysis
\end{abstract}
\begingroup
\renewcommand\thefootnote{\relax}
\footnotetext{This work is licensed under CC BY-NC-ND \url{http://creativecommons.org/licenses/by-nc-nd/4.0/}.}
\endgroup
\section{Introduction}
Heat dissipation is an important bottleneck in development of modern electronics circuits, namely when increasing their computational power together with shrinking the size. To circumvent this problem, various active or passive thermal management methods are being developed \cite{Dhumal2023, Chen22}. A key material property to design heat management structures is thermal conductivity, which reflects the ability to transfer the heat from hot to cold areas on the device \cite{Tu24, Pop23}.
Scanning Thermal Microscopy (SThM) \cite{Gomes15} is a Scanning Probe Microscopy (SPM) technique allowing measurement of local temperature, thermal conductivity and local thermomechanical properties. In the most common variant, it is based on the use of an SPM probe equipped by a resistive heating and/or temperature sensing element at its apex. The force between the probe and sample can be measured and adjusted, and the sample can be scanned using commonly used SPM methods. A dedicated circuit, typically a Wheatstone bridge, is used to measure the probe temperature or to heat it locally, depending on the measurement scenario. Here we consider the local thermal conductivity measurement, where the probe is heated and heat losses, which are proportional to the sample thermal conductivity, are monitored. Being based on SPM technology, SThM is unique in its capabilities to perform analysis of micro- and nanostructures with high spatial resolution \cite{Puyoo11, Saci14, Juszczyk}.

When the local thermal conductivity is measured, there are too many unknowns to evaluate the quantity directly. The probe-sample contact resistance and various parasitic heat flow paths are too complex to be measured or theoretically evaluated, and therefore a calibration of the whole system using samples with known thermal conductivity is the most viable approach for practical use of the method. The calibration samples can consist either of bulk material \cite{fischer} or of thin films on a substrate, mimicking various bulk thermal conductivities by averaged film-substrate bulk properties \cite{Guen20}. The measurement protocol is then based on the calibration of the complete instrument including the probe and electronics for its particular settings, i.e. the current flowing  through the probe and the bridge balance resistor values. The calibration curve is constructed from the SThM data obtained from the calibration samples and is used for the interpretation of data obtained from unknown samples. It is hereby assumed that neither the settings nor any other influence factor, such as the probe apex, changes significantly.

Practically, SThM calibration is a challenge from both experimental and data processing aspects. 
The calibration samples must fulfill several requirements. Firstly, they must have minimum surface roughness, in order to keep changes in the contact between the probe and the sample as low as possible. Secondly, they should possess good time stability, without e.g. oxidation. Thirdly, they should be traceable themselves, i.e. be suitable for some other traceable thermal conductivity measurement method. Lastly, they should be small enough to be used in the microscope chamber with minimal sample exchanges which require thermal stabilization and thus increase the measurement time. 
The range of potential bulk samples satisfying all of these requirements is limited. 
Available samples tend to have thermal conductivities clustered in certain regions, e.g. polymers (around 0.1 \WmK), glass and oxides (around 1 \WmK), metals (more than 10 \WmK).
Moreover, in order to minimize the impact of the, unavoidable, drifts in the electronics, the measurement protocol needs to include repeated measurement at a position far enough from the sample or on some reference sample (e.g. one of the calibration samples) which further limits the number of bulk samples that can be measured in a reasonable time. 
The time constraint is important not only for practical considerations but also because of potential issues with the stability of the measuring system, notably the probe and electronics.
Larger amount of measured samples or larger number of repetitions provide better calibration results, however increase the risk of damaging the probe.
In the end, the calibration curve, while being quite non-linear, has to be setup from a few points only and these points can have relatively large uncertainties. To propagate the uncertainty through the calibration process is therefore not trivial and an advanced statistical treatment is needed to get maximum of information from the available data.

We build on a paper by Fleurence et al. \cite{fleurence2023} that used Bayesian statistical inversion for the joint estimation of the parameters of the calibration curve as well as the thermal conductivities of the samples of interest.  
This approach is highly commendable for several reasons. Firstly, it provides a comprehensive treatment of uncertainties, accounting for errors in both calibration measurements and the thermal conductivity values of reference samples. By integrating uncertainties into both the calibration and prediction processes, it ensures traceable and reliable measurements. Secondly, the simultaneous estimation of the calibration parameters and thermal conductivities of the samples of interest enables an integrated approach that avoids biases introduced by separating these tasks. Thirdly, the Bayesian framework is particularly effective in handling non-linear and implicit calibration models, such as the one describing the relationship between SThM measurements and thermal conductivity. In addition, it provides probabilistic outputs in the form of posterior distributions, which are invaluable for robust uncertainty quantification and credible interval estimation. 

Despite these advantages, the method has certain limitations. Bayesian inversion can be computationally intensive, especially when applied to complex models requiring iterative sampling methods like Markov Chain Monte Carlo (MCMC). Moreover, while the approach offers detailed uncertainty analysis, it relies heavily on the quality and availability of prior information, which, if not carefully chosen, can introduce biases. Another limitation is its reliance on manual or semi-automated data processing steps, which may not scale effectively for large datasets or routine applications.

To address these challenges while building on the strengths of the Bayesian approach, we propose a new method based on the OEFPIL algorithm (Optimum Estimate of Function Parameters by Iterated Linearization). This algorithm is specifically designed for fitting nonlinear errors-in-variables regression models with potentially nonlinear constraints on their parameters, even in cases where the model variables are correlated. OEFPIL is computationally efficient and particularly well suited for calibration models. 

Unlike the Bayesian approach, which relies on probabilistic inference, OEFPIL operates within a deterministic framework, making it faster and less dependent on prior knowledge about the distribution of model parameters. Furthermore, our method incorporates a fully automated process for deriving calibration curve data directly from SThM images. This automation not only simplifies and accelerates the calibration process, but also represents a significant step toward making SThM calibration more practical, scalable, and widely applicable.

OEFPIL can be used for any measurement model involving both directly and indirectly measured quantities, a concept introduced by Kubáček in \cite{kubacek-foundations}. Indirectly measured quantities can be, e.g. function parameters or the unknown thermal conductivities as in this case. 

We demonstrate the presented method on a calibration of a new, physically smaller, set of bulk calibration samples, designed to be used within a vacuum SThM which has even tighter sample size limitations than a normal microscope running in ambient conditions. The goal of experiment used for OEFPIL demonstration in this paper was to use the existing larger calibration samples from Quantiheat FP7 project \cite{Quantiheat} to provide traceability for the new set of smaller and compactly packed samples.

\section{Methods}
\subsection{Measurement model}\label{sec:meas_model}
As described in the following experimental section, SThM measurements are based on the measurement of the changes in the temperature of the heated resistive probe as it gets cooled by the contact with the sample.
The temperature change is measured using a Wheatstone bridge, which is a circuit which is very sensitive to small changes in resistance. An output of the bridge is the voltage, which is proportional to the resistance change. To reduce the impact of electrical and temperature drifts and the impact of heat flowing back to the cantilever, the value measured on the sample is typically compared with a reference value, e.g. in air, far from the sample, or, as in our case, on a single reference sample. The primary quantity to process further is then the difference or ratio between the voltage on the studied sample and the one obtained on this reference measurement.

We use the generally used relationships between heat flux to the thermal conductivity, for a theoretical background see Fleurence et. al. \cite{fleurence2023} and references therein. However, we modify their definition of the intermediate measurand $Y$ to accomodate the differences in the setup and define it as the difference between the voltage measured on the sample and the voltage measured on the reference sample SiO$_2$ 
\begin{equation}
Y = U_{\mathrm{sample}}-U_{\mathrm{SiO}_2}.
\end{equation}
Thus we can work with a calibration curve of the same shape as in \cite{fleurence2023}
\begin{equation}
    Y = \frac{ak}{b+k} +c. \label{eq:calcurve}
\end{equation}
Here $a$, $b$ and $c$ are unknown parameters which describe the thermal properties of the probe sample contact as well as the interaction with the cantilever and environment.

\subsection{Experimental} \label{sec:experimental}
Measurements were performed on a Bruker Dimension Icon instrument. For SThM operation we have used our custom-built electronic system based on a Wheatston bridge. All the measurements were performed using VITA-DM-GLA1 thermal probes sold by Bruker and manufactured by Kelvin Nanotechnology.

The electronic system was designed to measure weak signals that are highly susceptible to noise, interference, and drift. The SThM probe is integrated into a Wheatstone bridge configuration, with its output amplified using the low-noise instrumentation amplifier AD8429. The bridge is excited by a precision voltage reference, the LT6657, which features very low temperature drift (1.5 ppm/°C) and excellent stability. This known voltage is a critical parameter for calculations related to the circuit's overall sensitivity.
The Wheatstone bridge itself is inherently sensitive to changes in one of its resistors. Any such change affects the current load of the bridge, which in turn impacts the excitation voltage and consequently alters sensitivity. To address this issue, additional test points were incorporated into the circuit to allow voltage measurements at critical nodes. Testing and analysis confirmed that the amplifier exhibits sufficient linearity and that the output voltage is directly proportional to the resistance changes in the SThM probe.
To minimize interference from mains power lines, the amplifier is powered by batteries. This approach also ensures galvanic isolation, eliminating potential ground loops and avoiding issues that might arise when measuring electrically conductive samples. The power supply is symmetrical, utilizing four lithium 18650 cells connected in series. Each voltage branch -- positive and negative -- is stabilized using ultra-low-noise voltage regulators, the LT3045 and LT3094, respectively. The output voltage is maintained at a fixed value of ±5\,V, regardless of gradual battery discharge.
The sensitive components of the circuit are enclosed in a metal shielding box. To reduce thermal interference, the batteries and stabilizers are placed separately from the sensitive parts. The SThM probe is connected via a double-shielded coaxial cable. Measurements are performed only after the system has reached thermal equilibrium to minimize parasitic Seebeck voltages. The circuit operates continuously, with the battery capacity sufficient for several days of operation. All resistors were selected for their low thermal sensitivity (10~ppm/°C), with the exception of potentiometers. However, potential drift in the potentiometers is monitored using the test points.
This electronic setup has proven effective for measuring subtle changes in the resistance of the SThM probe. When a fixed resistor replaces the probe, the system exhibits significantly lower noise and improved stability, indicating that residual noise and instability are primarily caused by air currents and temperature variations around the probe. 

Bulk samples from the Quantiheat project were used for SThM calibration. 
Samples were measured using the laser flash method by the French National Metrology Institute (LNE) during the Quantiheat project, and in this study, the LNE values and uncertainties were used as a reference.  
The calibration sample set consisted of six reference samples: PMMA (polymethyl methacrylate), POM-C (polyoxymethylene copolymer), SiO$_2$ (silicon dioxide), glass, Al$_2$O$_3$ (aluminum oxide), and p-doped silicon (silicon doped with positive charge carriers). The number of the calibration samples was chosen to keep the time requirements within one day of (automated) measurement.

The unknown set, to be calibrated, consisted of five bulk samples bought from Thorlabs, originally designed as 12 mm diameter, 3 mm thick optical flat windows, cut to smaller pieces and glued on a sample holder together with a 3 mm thick molded acrylic lens. 
In this manner, a sample with a diameter of less than 12 mm was prepared, consisting of five bulk materials, designated as samples A through E. These materials cover a wide range of nominal thermal conductivities: BK7 glass (A), fused silica (B), calcium fluoride, CaF$_2$ (C), germanium Ge (D) and acrylic glass (E).

The measurement protocol followed the work by Fleurence et al. \cite{fleurence2023} with a few deviations, allowing to make the process automated and better suited for unknown samples. 
\begin{enumerate}
	\item Measure a $5 \times 5$~$\mu$m$^2$ image on the first sample (either calibration sample or to-be-measured sample).
	\item Measure a $5 \times 5$~$\mu$m$^2$ image on the reference sample. 
	\item Repeat steps 1 and 2 four more times for the selected sample
	\item Repeat steps 1--3 for the remaining calibration and to-be-measured samples with the same reference sample.
\end{enumerate}
The original protocol was based on acquiring sets of single point measurements, where the probe touched the sample surface without laser feedback (dark mode). This approach was used to minimize the uncertainties related to feedback laser impact on the thermal data. However, an assumption about the local sample smoothness needed to be done, which might be suitable for good calibration samples, however might not be fulfilled by the unknown samples that we want to calibrate. In order to increase the amount of data obtained on a potentially rough sample and to address a wider range of probe-sample contact geometries that can be expected on such samples, we have measured full images (of size $5 \times 5$~$\mu$m$^2$), even if we had to use the laser feedback which could increase our uncertainty. The use of the feedback also allowed us to perform the measurements of all the samples automatically, including a reference measurement on a selected bulk sample (here SiO$_2$) between every other measurements in order to minimize impact of drifts due to temperature changes in the laboratory or coming from the electronics. The whole calibration procedure was programmed on our Dimension Icon microscope using its Programmed move tool. All the samples were placed on a large stepper motor equipped by a sample stage which was used for automated measurements. As the programmed move and acquire script had no option which would allow setting up the waiting time for stabilization between individual sample measurements (as suggested by the protocol in Ref. \cite{fleurence2023}) we performed the measurement without waiting, but used only the second half of the data for the calibration (i.e. $5 \times 2.5$~$\mu$m$^2$), throwing away the part of data measured directly after approach and establishment of feedback. Programmed mode also does not allow measurement in air, which was one of the reasons why SiO$_2$ was measured as a reference between the measurements. The complete procedure for measuring 6 calibration and 5 unknown samples took approximately 10 hours.

\subsection{Data processing}
For each data image we calculated the mean value and its standard deviation from the second half of the image, with an area of $5\times 2.5$~$\mu$m$^2$ and a resolution of 64 $\times$ 32 pixels. For each image on a measured sample, both unknown and calibration, we computed the difference between the voltage measured on the sample and the average voltage on the reference SiO$_2$ measured before and after. We estimated the uncertainty corresponding to a different landing/location using the Mandel-Paule estimator \cite{MandelPaule} obtaining thus a single value for each sample.  
The Mandel-Paule estimator is a widely used method in meta-analysis and interlaboratory studies to estimate a common mean or consensus value from data involving multiple sources of variability. It is particularly effective in cases where the measurements are provided with varying levels of uncertainty. This method accounts for both within-group variability and between-group differences, offering a robust approach to determining a consensus value. 

In our application, the Mandel-Paule estimator is used to assess a common value based on measurements taken from different locations on a sample. The corresponding uncertainty estimate reflects both the inherent inhomogeneity of the sample and the deviations between measurements while incorporating the varying uncertainties of the individual measurements. Based on the standard deviation of the mean, the estimator captures the effect of local roughness, providing a comprehensive evaluation of the variability within the sample.

The overall data processing thus consists of the following three steps:
\begin{enumerate}
	\item For each AFM image, determine the mean voltage measured during the second half of the image and its uncertainty. If necessary, apply a filter to remove scars, contaminations, etc. If sudden a jump occurred, split the data into two groups. 
	\item Determine difference between the voltage measured on the sample and the voltage on the reference sample measured before and after this sample. If a jump occurred, use only one-sided difference with the appropriate reference sample value.
	\item Combine all values obtained for a given sample using the Mandel-Paule estimator to obtain the intermediate measurand $Y$.
\end{enumerate}

\subsection{Data fitting}\label{sec:data_fitting}
The OEFPIL algorithm, originally developed for data fitting, can also be applied to any linear model with constraints. A key advantage of the used modeling approach, which is based on errors-in-variables regression, is its ability to distinguish between direct and indirect measurands -- a concept defined by Kubáček \cite{kubacek-foundations} and further refined in subsequent works. For additional details, see, e.g., \cite{oefpil-clanek, campbell2025, witkovsky2025}.

In many calibration scenarios, the calibration curve is expected to remain valid over an extended period. This approach makes it more practical to initially determine the calibration curve and then, as needed, use it repeatedly to estimate values for unknown samples.
However, in SThM, the calibration curve depends essentially on the tip apex, which is prone to wear, and on the settings of the electronics. Therefore, the calibration curve can be only rarely reused and in most cases we can determine all values together.

As explained in section \ref{sec:meas_model} we fit the data using the calibration curve \eqref{eq:calcurve} which relates the true values of the measurand \(Y\), say \(Y^*\), to the true value of the thermal conductivity \(k\), say \(k^*\). The relationship between the true values is expressed by equation:
\[
Y^* = \frac{ak^*}{b + k^*} + c,
\]
where \(a\), \(b\), and \(c\) are fitting parameters.

Here, we adopt the notation from \cite{witkovsky2025}, which facilitates straightforward implementation of the OEFPIL algorithm. For that, let the true values of the directly measured quantities be denoted by \(\mu_1, \dots,\mu_N\), \(\mu_{N+1},\dots,\mu_{2N}\), and  \(\mu_{2N+1}, \dots \mu_{2N+M}\), and let the indirect measurands, the unknown model parameters of main interest, be denoted by \(\beta_1, \dots, \beta_P\), and \(\beta_{P+1}, \dots, \beta_{P+M}\). 

Although the quantities \(\mu_1, \dots, \mu_{2N+M}\) are generally unknown, they are observable and can be estimated through direct measurements. Specifically, \(\mu_1, \dots, \mu_{N}\) represent the true values of the conductivities of the \(N\) reference samples, \(k^*_1, \dots, k^*_N\); \(\mu_{N+1}, \dots, \mu_{2N}\) represent the true values of the \(Y\)-measurands, \(Y^*_1, \dots, Y^*_N\), for the \(N\) reference samples; and \(\mu_{2N+1}, \dots, \mu_{2N+M}\) represent the true values of the \(Y\)-measurands, \(Y^*_{N+1}, \dots, Y^*_{N+M}\), for the \(M\) new (unknown) samples. 

Furthermore, the indirect measurands \(\beta_1, \dots, \beta_P, \beta_{P+1}, \dots, \beta_{P+M}\) include \((\beta_1, \dots, \beta_P)\), the \(P\) calibration curve parameters, and \((\beta_{P+1}, \dots, \beta_{P+M})\), the \(M\) thermal conductivities of the unknown samples.

Using this notation, and assuming \(N \geq 3\), we find that the \(q = N + M\) constraints on the unknown model parameters can be expressed as:
\begin{eqnarray}
    \frac{\beta_1 \mu_i}{\beta_2 + \mu_i} + \beta_3 - \mu_{N+i} &=& 0, \qquad i = 1, \dots, N, \label{eq:constraint_A} \\
    \frac{\beta_1 \beta_{3+j}}{\beta_2 + \beta_{3+j}} + \beta_3 - \mu_{2N+j} &=& 0, \qquad j = 1, \dots, M. \label{eq:constraint_B}
\end{eqnarray}
With \(N = 6\), we take \((\mu_1, \dots, \mu_6) = (k^*_1, \dots, k^*_6)\), representing the thermal conductivities of the six reference samples, and \((\mu_7, \dots, \mu_{12}) = (Y^*_1, \dots, Y^*_6)\), representing the true values of the measurands \(Y\) for the six reference samples. With \(P = 3\), we take \((\beta_1, \beta_2, \beta_3) = (a, b, c)\), representing the calibration curve parameters, and with \(M = 5\), we have \((\beta_{4}, \dots, \beta_{8}) = (k^*_{A}, \dots, k^*_{E})\), representing the thermal conductivities of samples A through E. Additionally, \((\mu_{13}, \dots, \mu_{17}) = (Y^*_A, \dots, Y^*_E)\) represent the true values of the measurands \(Y\) for samples A through E.

The conventional approach, to fit only the curve parameters $a$, $b$, $c$ and determine the thermal conductivities of the unknown samples from the calibration curve, is also possible within this framework. In that case, we work only with the constraints \eqref{eq:constraint_A} reducing the $\mu$ parameters to $\mu_1, \dots, \mu_{2N}$ and the $\beta$ parameters to $\beta_1, \beta_2, \beta_3$.

The fitting was performed using the Matlab implementation of OEFPIL, which directly provides the estimates of all \(\mu\) and \(\beta\) parameters, along with the corresponding covariance matrix.

\section{Results}
Two complete, independent sets (dataset 1 and dataset 2) of measurements were performed. In both cases, 6 reference samples and 5 unknown samples were measured using the protocol described in section \ref{sec:experimental}. Out of the 5 unknown samples, one sample had to be removed from the dataset 2 as results were inconsistent with the other measurements and the assumed model. 

\begin{figure}[htbp]
	\centering
	\begin{tabular}{cc}
		\includegraphics[width=0.5\textwidth]{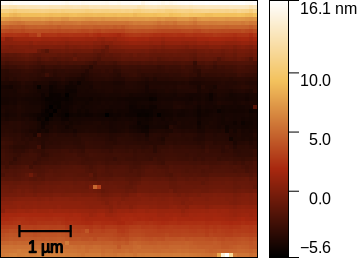} &
		\includegraphics[width=0.5\textwidth]{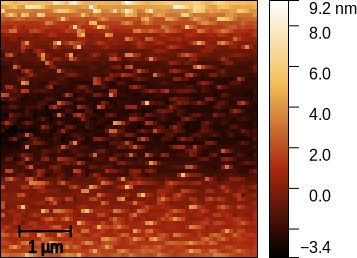} \\
		\includegraphics[width=0.5\textwidth]{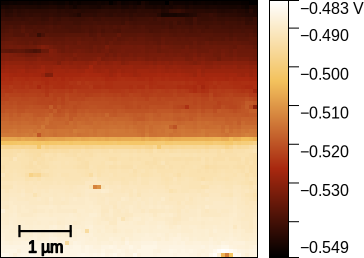} &
		\includegraphics[width=0.5\textwidth]{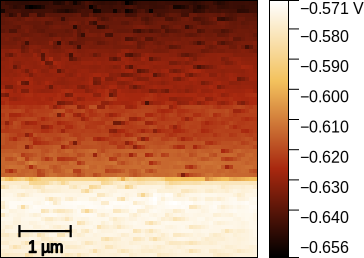} 
	\end{tabular}
	\caption{Examples of problematic SThM images of topography (top) and voltage (bottom) for two samples -- germanium (left) and ZrO$_2$ (right). On the left, the jump is clearly related to the change in the topographic signal, on the right the change is only in the voltage signal.}
	\label{fig:afm-jump}
\end{figure}

In some cases, the probe underwent a sudden change of state resulting in a jump in the voltage value measured, see for illustration Fig.~\ref{fig:afm-jump}. This is often, but not always, related to a sudden change in the measured topography due to contamination or tip apex change. These cases cannot be processed routinely but require either manual treatment or more sophisticated algorithms. In these cases it is not possible to subtract the voltage from the average voltage of the reference sample measured before and after the problematic image but the value acquired on the reference sample before or after the jump should be used, as appropriate. In order to evaluate the impact of these effects we performed the corrections, including removal of point defects, manually for  measurement set 1, resulting in dataset 3. Problematic measurements occured more often on the unknown samples (13 out of 25 measurements) than on calibration samples (8 out of 30 measurements) or the SiO$_2$ reference sample (8 out of 56 measurements, often when alternating with a problematic sample). 
In terms of the measurand $Y$, this lead to a decrease of the overall uncertainty for measurements with a significant amount of defects in form of outliers. For samples with discontinuous jumps the improved processing correctly increased the uncertainty. The changes were up to 20~\% in terms of the value and up to 60~\% in terms of the uncertainty. 

\begin{figure}[htbp]
	\centering
	\begin{tabular}{cc}
		\includegraphics[width=0.5\textwidth]{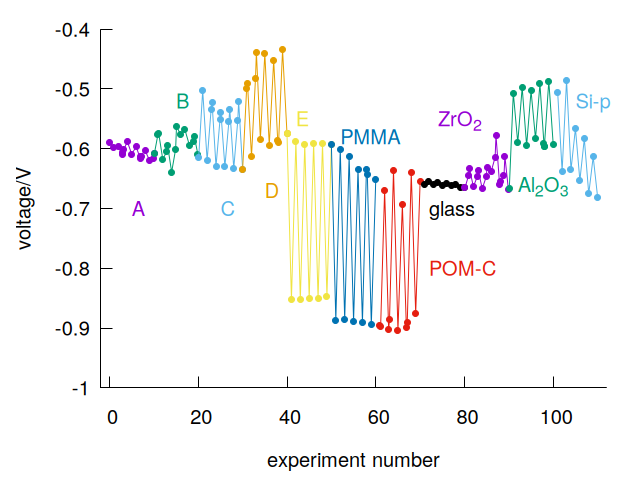} &
		\includegraphics[width=0.5\textwidth]{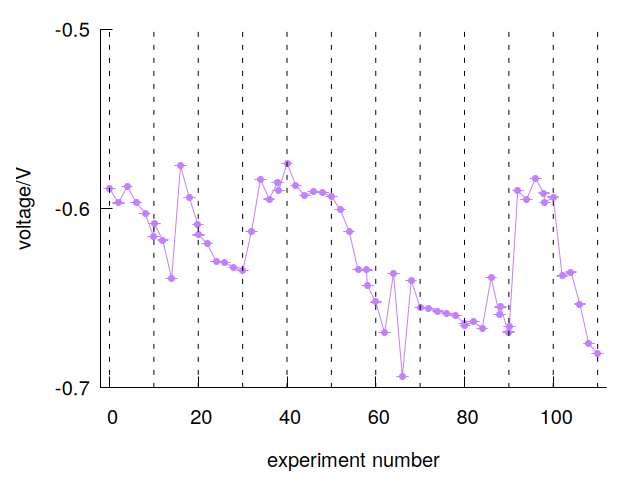} 
	\end{tabular}
		\caption{Left: Example of the voltage signal aquired within dataset 3. Uncertainties are too small to be visualized. The data are affected by noticeable drift. Right: Evolution of voltage measured on the reference sample SiO$_2$. The dashed lines denote the changes of the sample with which the reference sample is alternated. Uncertainties are mostly below 10$^{-4}$ and cannot be distinguished here. }
	\label{fig:voltage}
\end{figure}

The voltage values are affected by long term drifts as well as sudden changes, as illustrated in the left part of fig.~\ref{fig:voltage}. On the right of Fig.~\ref{fig:voltage} the evolution of the voltage on the reference sample SiO$_2$, which in theory should be constant, is highly nontrivial. This does not take into account the actual time dependence, since the measurements are not equally spaced in time. There is no option to set the time precisely within the framework of the microscope and the paths the head of the microscope must travel between samples varies.
In order to minimize the effect of the drift  we form the voltage difference from the values of three consecutive measurements and process these voltage differences separately. Combining the values from five measurements per sample combines the variability due to sample inhomogeneity as well as the effect of drift. As noted in Fig.~\ref{fig:voltage}, the uncertainties due to noise within an image are much smaller than the effect of the drift.
The drawback of this method is the inability to distinguish between the inhomogeneity of the sample and the drift. In order to clarify this it would be necessary to perform multiple measurements at the same location, thus keeping only the drift involved. This however would set only a lower limit, since in reality the drift may be much more complex. Sudden changes can be related to changes of the tip apex which in turn depend on the sample and its topography.
In terms of the measurand $Y$, the uncertainties of the individual terms are around $2 \times 10^{-4}$, whereas the combined uncertainty for the whole sample is around $10^{-3}$.


We can combine dataset 1 and dataset 2 using the Mandel-Paule estimator, resulting in dataset 4. Note however, that estimating an unknown source of uncertainty related to different datasets based on two sets only is not very reliable. More repeated measurements would be necessary, requiring approximately two weeks of measurement time in total.

The two procedures outlined in Section \ref{sec:data_fitting}, namely, (i) fitting the calibration curve simultaneously with the unknown thermal conductivities and (ii) fitting only the calibration curve parameters in the first step of the two-step procedure and then predicting the unknown thermal conductivities in the second step, produce identical results in terms of the estimated curve parameters \(a\), \(b\), \(c\), the estimated thermal conductivities \(k^*_A, \dots, k^*_E\), and their uncertainties. 
This equivalence is based on two main reasons. First, it arises from the simple form of the constraints defined in \eqref{eq:constraint_B}, which  effectively expresses the inversion of the calibration curve at given $Y$'s. Furthermore, the random variables associated with the estimator of the parameters  \(\beta\) (the calibration curve parameters) and the estimators of \( (Y^*_A, \dots, Y^*_E)\) (the true values of the measurands \(Y\) for samples A through E) are uncorrelated, as the estimates of \(\beta\) do not depend on \( (Y^*_A, \dots, Y^*_E)\).
In the first step of the two-step procedure, only the calibration curve parameters are estimated. These estimates, along with their uncertainty matrices, are then used in the second step to predict the thermal conductivities of new, unknown samples using the measured values of \( (Y^*_A, \dots, Y^*_E)\) and their associated uncertainty matrices. As noted, the estimators of the calibration curve parameters does not depend on the true values \( (Y^*_A, \dots, Y^*_E)\) or their estimators. Consequently, their joint uncertainty matrix is block-diagonal, and here, in this particular case it is even diagonal. This property of separability ensures that the results obtained from the two-step procedure are equivalent to those from the one-step approach. The situation would differ if the uncertainty matrix for direct measurements were not block-diagonal.

\begin{figure}
    \centering
    \includegraphics[width=0.75\linewidth]{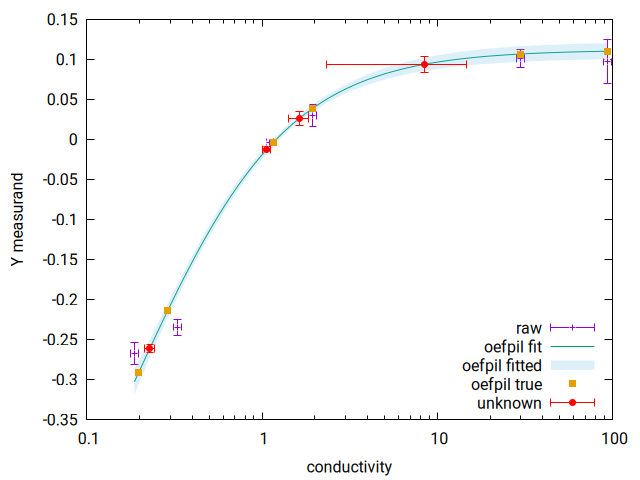}
	\caption{Calibration curve for dataset~3. The uncertainties shown in the image are expanded.}
    \label{fig:cal-curve-C}
\end{figure}

An example of the resulting calibration curve is shown in fig.~\ref{fig:cal-curve-C}. We can clearly see that for high thermal conductivities the function changes only little, leading to high uncertainties in conductivity. The method will be the most useful in the range of the steep slope between approx. 0.1 and 10 \WmK. 

A comparison of the values obtained from the datasets 1--4 is shown in table~\ref{tab:cal-ABCD}. As can be seen therein, the precise values of the calibration curve parameters can differ significantly between different measurements. This is commented on in a following paragraph. 
The resulting estimates for the unknown conductivities agree within uncertainties, although the values obtained from dataset~2 are all higher than their counterparts from dataset~1. The manual postprocessing performed on dataset~1 resulting in dataset~3 led to a reduction in the uncertainties of the thermal conductivities between 15 and 60 \%, underlining the need of correct processing of data biased by contamination etc. Also note that the correlation between calibration curve parameters is very high, it is fairly low for the unknown conductivities. In all cases, the fit exhibit very low p-values.  This indicates either a problem with the assumed calibration curve expression or significantly underestimated uncertainties. However, the source of these uncertainties and thus their quantification is unclear so far.

For sample D the \(Y\)-measurand was higher than the asymptotic value for large thermal conductivities in both measured datasets 1 and 2, for sample C in datasets 2 and 4. In these cases no meaningful value for the thermal conductivity could be deduced. The problem is probably due to underestimated uncertainties resulting in a too narrow uncertainty band of the calibration curve. The calibration curve is almost flat for high thermal conductivities as shown in Fig.~\ref{fig:cal-curve-C} which suggests that the method will not give useful results for thermal conductivities above approx. 10 \WmK. This corresponds to the huge uncertainties found for sample C from datasets 1 and 3, which make this method practically useless for such a sample. The presence of such high \(Y\)-measurand values may affect the convergence of the fitting process, if we fit both curve parameters and thermal conductivities. In fact, in some cases convergence could not be achieved. On the other hand, if convergence has been achieved then the results obtained from data including the problematic point and those without do not differ.  

\begin{figure}[htbp]
	\centering
	\includegraphics[width=0.75\linewidth]{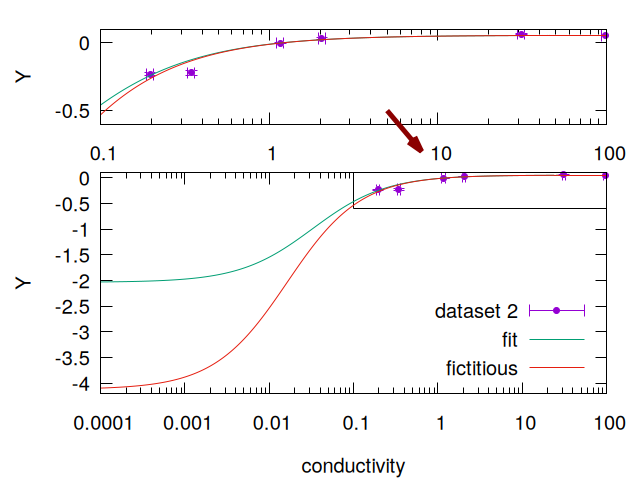}
	\caption{The values of the calibration samples shown in the top part, cover, in reality, only a small part of the range of the calibration curve (bottom). The fictitious curve is barely distinguishable in the top image but corresponds to a very different combination of parameter values, notably the combination ($2a$, $b/2$, $c-a$) instead of the original combination $(a,b,c)$. The difference becomes noticeable if the full range is viewed (bottom). }
		\label{fig:uncabc}
\end{figure}

It is important to emphasize that the values of the calibration curve parameters $a$, $b$, $c$ themselves do not have any direct physical meaning. The available calibration samples cover only a small range of the calibration curve and, as illustrated in Fig.~\ref{fig:uncabc}, different combinations of parameters may give rise to very similar curves. This may pose a problem if the calibration samples are measured with high uncertainties or don't lie close enough to the curve.  

The uncertainties of thermal conductivities determined from dataset~2 are higher due to the higher uncertainties of the \(Y\)-measurand and the positions shifting closer to the high conductivity asymptote.

\begin{table}[htb]
    \centering
    \addtolength{\tabcolsep}{-0.25em}
        \begin{tabular}{ccccc}
	Dataset	& 1 & 2 & 3 &4 \\ \hline
		$a$ & 0.770 $\pm$ 0.061 
& 2.09 $\pm $0.64
		& 0.825 $\pm$ 0.057 
		&1.18 $ \pm$ 0.23 \\ 
     $b$ & 0.200$\pm$ 0.035 & 0.032 $\pm$ 0.011  
		& 0.189$\pm$ 0.024 
		&0.102 $ \pm$ 0.028\\
     $c$ & -0.661 $\pm$ 0.067  &
     -2.03 $\pm$ 0.63 
		& -0.713 $\pm$ 0.060
		& -1.09 $ \pm$ 0.23\\
     $r_{a,b}$ & -0.951 & -0.998  
		& -0.956 
		& -0.991 \\
     $r_{a,c}$ & -0.997 & -0.999  
		& -0.98 
		& -0.999\\
     $r_{b,c}$ & 0.971 & 0.998  
		& 0.971 
		& 0.992\\\hline 
     $k_A$ & 1.047  $\pm$ 0.034  &1.23  $\pm$ 0.18 
		& 1.065 $\pm$ 0.029 
		& 1.118 $ \pm$ 0.068 \\ 
     $k_B$ & 1.60   $\pm$ 0.26 &3.6 $\pm$ 2.1 
		& 1.63 $\pm$ 0.11
		& 1.91 $ \pm$0.15 \\ 
     $k_C$ & 9.0   $\pm$ 4.4 &  ---  
		& 8.5 $\pm$ 3.1 
		&  ---\\ 
     $k_E$ & 0.2132 $\pm$ 0.0086 & 0.2174  $\pm$ 0.0060  
		& 0.2290 $\pm$ 0.0072 
		& 0.263 $ \pm$ 0.019
	\end{tabular}
	\caption{Comparison of the calibration curve parameters $a$, $b$ and $c$, their correlation coefficients $r$  and of the conductivities of the unknown samples as determined from the datasets 1--4 (1 and 2 are measured, 3 is a manual correction of 2, 4 is the common mean of 1 and 2). Uncertainties $u$ are not expanded, correlation coefficients are given as well. Sample D gave rise to too high values of the \(Y\)-measurand, which were not compatible with the model. }
    \label{tab:cal-ABCD}
\end{table}

The slight flattening of the calibration curve between the datasets~1 and 2 may be a sign of a change in the thermal conductance of the probe sample contact, but more measurements would be needed to confirm this. 

\begin{figure}
    \centering
    \begin{tabular}{cc}
    \includegraphics[width=0.45\linewidth]{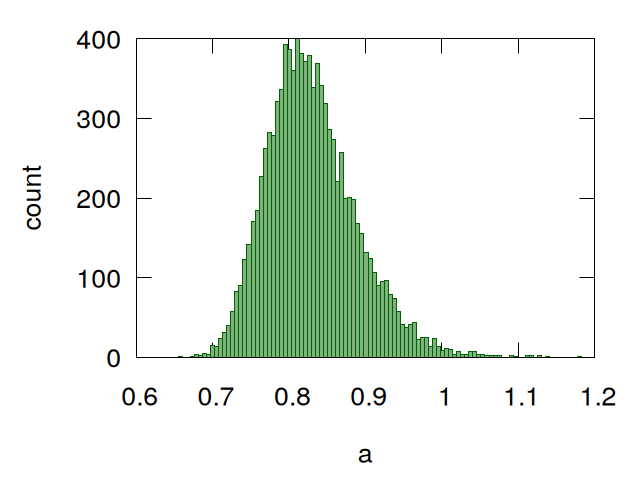} &
    \includegraphics[width=0.45\linewidth]{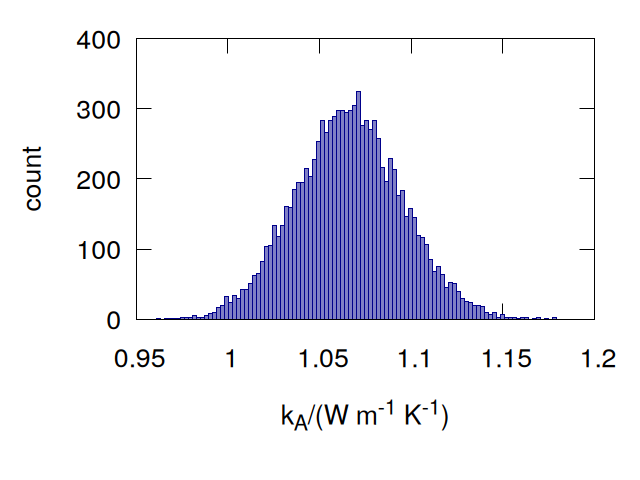} \\
    \includegraphics[width=0.45\linewidth]{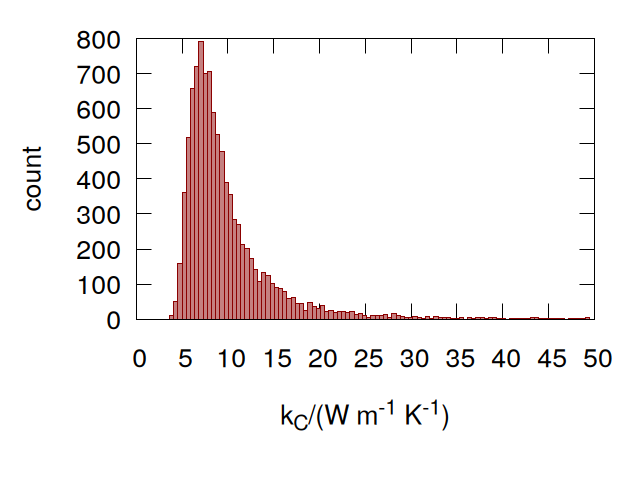}
    \end{tabular}
    \caption{Distributions of the calibration curve parameter $a$ (top left) and the thermal conductivity for samples A (top right) and C (bottom) obtained from dataset 3 using Monte Carlo.}
    \label{fig:MC-pdf}
\end{figure}

The results were compared to values obtained by Monte Carlo. Both procedures, i.e. either fitting only curve parameters or fitting both curve parameters and conductivities, were performed. We assume normal distributions of the thermal conductivities of the calibration sample as well as of the \(Y\)-measurands of calibration and measured samples. We do not assume any correlation. 
Noise generated according to the appropriate distributions has been added to the estimated true values. The algorithm did not converge for all generated datasets. This was especially problematic for dataset 2. The results are given in Table~\ref{tab:MC} typical probability distributions for a curve parameter, a thermal conductivity within the limits of SThM capabilities and a thermal conductivity at the limit is shown in Figure \ref{fig:MC-pdf}. The agreement is in general reasonable, with the exception of $k_C$ for dataset~1. This is probably due to the large asymmetry of its probability distribution, as seen in Figure \ref{fig:MC-pdf}. In general, the mean and the standard deviation cannot capture the behavior of an asymmetric distribution and a coverage interval should be used instead. In this case, the wide spread of possible values is another sign of the method reaching its limits for the sample.

\begin{table}[htb]
    \centering
    \begin{tabular}{c|c|cc}
	    Dataset & Quantity  &  OEFPIL & Monte Carlo  \\ \hline
	    \multirow{3}*{1}  &  a & $0.7700 \pm 0.0612 $ &  $0.7735 \pm 0.0596 $ \\
                         &  b & $ 0.1996\pm 0.0346 $ &  $0.2036 \pm 0.0326 $ \\
                         &  c & $-0.6611\pm  0.0673  $ &  $-0.6635 \pm 0.0650 $ \\ 
			 & k$_A$ &  $1.0474 \pm 0.0338$ & $1.0482 \pm 0.0334$\\
			 & k$_B$ &$ 1.599  \pm  0.262 $& $1.636\pm 0.293 $\\
			 & k$_C$ & $9.05 \pm  4.39 $& $27.3 \pm 1134.5 $ \\
			 & k$_E$ & $ 0.21324 \pm 0.00858$ & $0.21308 \pm 0.00868 $ \\
			 \hline
       \multirow{3}*{2}  &  a & $2.087\pm  0.635 $ & $2.793 \pm 1.309 $ \\
                         &  b & $ 0.0327\pm  0.0109  $ &  $-0.0056 \pm 0.4182
 $ \\
                         &  c & $ -2.032\pm 0.635  $ &  $-2.74 \pm 1.31$ \\ 
			 & k$_A$ & $  1.230 \pm  0.181 $ &  $1.233 \pm 0.188 $ \\ 
			 & k$_B$ & $  3.58 \pm 2.08  $ &  $ 4.01 \pm 3.91 $ \\ 
			 & k$_C$ & --- & --- \\ 
			 & k$_E$ & $  0.21740 \pm  0.00602 $ &  $0.236 \pm 0.327 $ \\ 
			 \hline
	    \multirow{3}*{3}  &  a & $ 0.8245\pm  0.0568 $ &  $0.8302 \pm 0.0588  $  \\
                         &  b & $ 0.1892\pm  0.0244 $ &  $0.1900 \pm 0.0241 $ \\
                         &  c & $-0.7128\pm  0.0601 $ &  $-0.718 \pm 0.0619 $ \\ 
			 & k$_A$ & $  1.0652\pm  0.0290 $ &  $1.0660 \pm 0.0291 $ \\ 
			 & k$_B$ & $  1.632 \pm  0.105$ &  $1.638 \pm 0.107 $ \\ 
			 & k$_C$ & $ 8.47 \pm 3.07 $ &  $11.7 \pm 40.2 $ \\ 
			 & k$_E$ & $  0.22901 \pm  0.00718 $ &  $0.22888 \pm 0.00715 $ \\ 
			 \hline
       \multirow{3}*{4}  &  a & $1.182 \pm  0.228  $ &  $1.243 \pm 0.331 $ \\
                         &  b & $ 0.10217\pm  0.0284 $ &  $0.1037 \pm 0.0282 $ \\
                         &  c & $-1.092 \pm  0.233 $ &  $-1.152 \pm 0.334 $ \\ 
			 & k$_A$ & $  1.1177 \pm  0.0680 $ &  $1.1208 \pm 0.0679 $ \\ 
			 & k$_B$ & $1.910\pm 0.155$  & $1.924 \pm 0.161 $ \\ 
			 & k$_C$ &  --- & --- \\ 
			 & k$_E$ & $0.2627 \pm  0.0194$ & 
             $0.2633 \pm 0.0197$ \\ 
			 \hline
    \end{tabular}
    \caption{Comparison of the curve parameters and thermal conducitivites including their uncertainties as obtained by estimating the curve parameters by OEFPIL and by the Monte Carlo method.}
    \label{tab:MC}
\end{table}

\section{Conclusions}
In this paper we presented an alternative procedure for the measurement of thermal conductivity using SThM including calibration. The main results are the following.
\begin{itemize} 
\item The proposed process is automated, which allows to significantly reduce the amount of time needed for the whole experiment, including calibration and measurement of unknown samples. The automation reduces not only the machine time but also labor costs, since the process does not have to be supervised constantly. This is especially important since calibration curves in SThM are not expected retain a long lifetime. 
\item The surface roughness or local contamination of the to-be-measured samples may be often problematic for SThM. In combination with an automated procedure, reliable data processing algorithms are needed to detect contaminations, drift, and other errors. Manual data processing of the data measured automatically is also possible; in most cases, it is probably still faster than the fully manual measurement. On the other hand, it is difficult to maintain an equivalent level of data processing (e.g. thresholds for filters) across the whole dataset during manual processing. Improved data processing has been shown to reduce uncertainty. In both cases, values, however, usually agree within error margins.
\item The algorithm OEFPIL was used to simultaneously determine the calibration curve and the values of the to-be-measured samples. Unlike the Bayesian approach it does not require any prior knowledge about the distribution of model parameters. As a standalone method, it has low demands for computational resources comparable to conventional fitting methods. The agreement with estimates obtained using the Monte Carlo method are approx a few percent for values and around ten percent for uncertainties.

\end{itemize}

Even if the automated calibration can in principle lead to higher final measurement uncertainty when compared to manual measurement sample by sample, it significantly reduces the necessary expertise and makes the SThM calibration procedure closer to daily laboratory practice. The use of the OEFPIL algorithm is another step in this direction, providing mathematically correct data fitting and uncertainty evaluation in a single step.

\section*{Acknowledgements}
The work was supported by the Czech Science Foundation project GA23­06263S and by the Czech Ministry of Education, Youth and Sports and the Slovak Research and Development Agency through the Inter-Excellence II program, project LUASK22008/SK-CZ-RD-21-0109.

\bibliographystyle{unsrt}
\bibliography{refSThM}

\end{document}